\documentclass{llncs}
\usepackage{llncsdoc}
\usepackage{graphicx}
\usepackage{subfigure}
\usepackage{multirow}
\usepackage{amssymb}
\usepackage{amsmath}
\usepackage{amstext}
\usepackage{amsgen}
\usepackage{textcomp}
\usepackage{amsbsy}
\usepackage{amsopn}
\usepackage{mathrsfs}
\usepackage{booktabs}
\usepackage{tikz}
\usepackage{balance}
\usepackage{hyperref}
\usepackage{upgreek}
\usepackage{enumitem}

\usepackage{todonotes}

\newcommand{\beq}{\begin{equation}}
\newcommand{\eeq}{\end{equation}}
\newcommand{\tr}[1]{{#1}^\top}
\renewcommand{\vec}[1]{\mathbf{#1}}

\newcommand{\mr}[1]{\mathrm{#1}}

\newcommand{\Real}{\mathbb{R}}

\newcommand{\GG}{\mathcal{G}}
\newcommand{\VV}{\mathcal{V}}
\newcommand{\EE}{\mathcal{E}}
\newcommand{\NN}{\mathcal{N}}
\newcommand{\LLambda}{\mathbf{\Lambda}}
\newcommand{\TTheta}{\Uptheta}
\newcommand{\LL}{\vec{L}}
\newcommand{\II}{\vec{I}}

\newcommand{\uhat}{\widehat{\vec{u}}}
\newcommand{\xx}{\vec{x}}
\newcommand{\yy}{\vec{y}}
\newcommand{\UU}{\vec{U}}

\newcommand{\Uhat}{\widehat{\vec{U}}}
\newcommand{\DD}{\vec{D}}
\newcommand{\AAA}{\vec{A}}
\newcommand{\mmu}{\boldsymbol{\mu}}
\newcommand{\ppm}{\,$\pm$\,}
\newcommand{\SSS}{S}
\newcommand{\DDD}{D}
\newcommand{\Lseg}{\mathcal{L}_{\mr{seg}}}
\newcommand{\Ldis}{\mathcal{L}_{\mr{dis}}}
\newcommand{\data}{\mathcal{X}}
\newcommand{\dataSrc}{\data_{\mr{src}}}
\newcommand{\dataTgt}{\data_{\mr{tgt}}}
\newcommand{\ypred}{\widehat{\yy}}
\newcommand{\zpred}{\widehat{z}}

\begin{document}

% For changing the color of the text

\newcommand*{\textred}{\textcolor{red}}
\newcommand*{\textblue}{\textcolor{blue}}
\newcommand*{\textgreen}{\textcolor{green}}

%% TENTATIVE TITLE:
\title{Graph Domain Adaptation for Alignment-Invariant Brain Surface Segmentation}
%\title{Adversarial Graph Convolutions training for Brain Surface Analysis}

%%%%%%%%%%%%%%%%%%%%%%%%%%%\title{Graph Convolutions on geometric aware Spectral representation: Learning Cortical Surface Data (?)} 

 \author{Karthik Gopinath \and Christian Desrosiers \and Herve Lombaert }
 \institute{ETS Montreal, Canada}
 
% \title{Graph Convolution on Spectral Embeddings: Application to Cortical Surface Parcellation} 

%  \author{************* }
%  \institute{***************}
% \author{Karthik Gopinath \and Christian Desrosiers \and Herve Lombaert }
% \institute{ETS Montreal, Canada}
% \email{karthik write your email id}}

\maketitle         

\begin{abstract}

The varying cortical geometry of the brain creates numerous challenges for its analysis. Recent developments have enabled learning surface data directly across multiple brain surfaces via graph convolutions on cortical data. However, current graph learning algorithms do fail when brain surface data are misaligned across subjects, thereby affecting their ability to deal with data from multiple domains. Adversarial training is widely used for domain adaptation to improve the segmentation performance across domains. In this paper, adversarial training is exploited to learn surface data across inconsistent graph alignments. This novel approach comprises a segmentator that uses a set of graph convolution layers to enable parcellation directly across brain surfaces in a source domain, and a discriminator that predicts a graph domain from segmentations. More precisely, the proposed adversarial network learns to generalize a parcellation across both, source and target domains. We demonstrate a \textbf{8\%} mean improvement in performance over a non-adversarial training strategy applied on multiple target domains extracted from MindBoggle, the largest publicly available manually-labeled brain surface dataset. 
\end{abstract}

\section{Introduction}

%In medical imaging applications, neuroimage analysis deals with understanding the working of the brain. The Magnetic Resonance Imaging (MRI) modality captures the anatomical and functional regions of the brain. 

The cerebral cortex is essential to a wide range of cognitive functions. Automated algorithms for brain surface analysis thus play an important role in understanding the structure and working of this complex organ. Nowadays, deep learning models such as convolutional neural networks (CNNs) have achieved state-of-the-art performance for most image analysis tasks including image classification, registration and segmentation \cite{arbabshirani2017single}. However, these models typically require large annotated data for training, which are often expensive to obtain in medical applications. This is especially true for the task of cortical segmentation, also known as \emph{parcellation}, where generating ground truth data requires labelling possibly thousands of nodes on a highly-convoluted surface. This burden also explains why datasets for such task are relatively small. For instance, the largest publicly-available dataset for cortical parcellation, MindBoggle \cite{Klein2017Mindboggling}, contains only 101 manually-annotated brain surfaces. Moreover, another common problem of deep learning models is their lack of robustness to differences in the distribution of training and test data. Hence, a CNN model trained on the data from a source domain usually fails to generalize to samples from other domains, i.e., the \emph{target} domains. 

Domain adaptation \cite{tajbakhsh2019embracing} has proven to be a powerful approach for making algorithms trained on source data to generalize on data from a target domain, without having explicit labels for target samples. Generative adversarial networks (GANs) \cite{goodfellow2014generative} leverage adversarial training to produce realistic images. In such approach, a discriminator network classifies images produced by a generator network as real or fake, and the generator improves by learning to fool the discriminator. Following the success of GANs, adversarial techniques have later been proposed to improve the learning capability of CNNs across different domains. In adversarial domain adaptation methods for segmentation \cite{zhang2017curriculum,zou2018unsupervised,ghafoorian2017transfer,vu2019advent,zhang2018task,javanmardi2018domain}, two main task are considered: the first one involves learning a fully supervised segmentator on the source domain and, in the second, a discriminator network forces the segmentator to have a similar prediction on both, the source and the target domains. These adversarial techniques usually rely on either feature space adaptation or output space adaptation. Initial works \cite{long2015learning,ganin2015unsupervised} focused on matching the distributions of features from source and target domains such that the learning generalizes across domains for classification tasks. As the output of CNNs for segmentation contains rich semantic information, \cite{tsai2018learning} proposed a method that, instead, leverages output space adaptation. Various work on pixel-wise domain adaptation has been developed for natural color images \cite{ganin2015unsupervised,hoffman2016fcns}. In medical image analysis, \cite{kamnitsas2017unsupervised} proposed an adversarial neural network for MRI image segmentation without any additional labels on the test domain. Likewise, \cite{javanmardi2018domain} presented a vessel segmentation approach for fundus images, which uses a gradient reversal layer for adversarial training. Recent work \cite{bateson2019constrained} also addressed the problem of domain adaptation by adding a differentiable penalty on the target domain. However, these domain adaptation techniques focus on data lying in the Euclidean space (natural or medical images), and are not suitable for graph structures such as surface meshes.

The image space is inadequate to capture the varying geometry of the brain surface. Differences in surface geometry hinder statistical frameworks from exploiting spatial information in Euclidean space. The extension of standard convolutions to non-Euclidean spaces like manifolds and graphs has led to the development of various geometric deep learning frameworks \cite{Bronstein2017Geometric,Monti2017Geometric}. A recent work \cite{cucurull2018convolutional} proposed to use geometric deep learning for segmenting three cortical regions by relying on the spatial representation of the brain mesh. Later, based on the spectral representation of brain meshes, \cite{gopinath2019graph} developed a graph convolution network (GCN) to parcellate the cerebral cortex. Despite offering more flexibility than Euclidean-based approaches, these methods are domain-dependent and would fail to generalize to new datasets (domains) without explicit re-training. Adding to the challenges of the field, obtaining annotations for these new datasets is also in practice particularly difficult. %\textred{cyclegan for cortical thickness?}

In this paper, we address the limitations of existing techniques for cortical parcellation and propose an adversarial domain adaptation method on surface graphs. Specifically, we focus on a problem shared by most GCN-based approaches, which is the need for a common basis to represent and operate on graphs. For instance, spectral GCNs \cite{bruna2013spectral,Defferrard2016Convolutional} require computing the eigendecomposition of the graph Laplacian matrix in order to embed graphs in a space defined by a fixed eigenbasis. As described in \cite{gopinath2019graph}, separate graphs may have different eigenbases. Furthermore, the eigenvectors obtained for a given graph are only defined up to a sign (i.e., $\pm1$), and up to rotation if different eigenvectors share close eigenvalues, typically observed in spectral graph analysis. Due to these ambiguities, spectral GCNs cannot be used to compare multiple graphs directly and need an explicit alignment of graph eigenbases as an additional pre-processing step, which brings its own ambiguities. Here, we focus on generalizing parcellation across multiple brain surface domains by removing the dependency to these domain alignments. 

The contributions of our work are multifold: 

\begin{itemize}[itemsep=3pt,topsep=3pt]
\item We present, to the best of our knowledge, the first adversarial graph domain adaptation method for surface segmentation. Our novel method trains two networks in an adversarial manner, a fully-convolutional GCN segmentator and a GCN domain discriminator, both of which operate on the spectral components of surface graphs.
\item Compared to existing approaches, our surface segmentation method offers greater robustness to differences in domain-specific alignments. Hence, our method offers a better generalization on target-domain datasets where surface data are aligned differently, without requiring an explicit alignment or manual annotations of these surfaces. 
\item We demonstrate the potential of our method for alignment-invariant parcellation of brain surfaces, using data from MindBoggle, the largest publicly-available manually-labeled surface dataset. Our results show a mean Dice improvement of \textbf{8\%} over using the same segmentation network without adversarial training. 
\end{itemize}

In the next section, we detail the fundamentals of our graph domain adaptation method for surface segmentation, followed by experiments validating the advantages of our method and a discussion of results.

%The distribution of samples in the dataset varies with samples collected from multiple clinical sites and acquisition protocols.

\section{Method}
\begin{figure*}[t!]
 \centering
%   \hspace{-15mm} 
 
    \includegraphics[width=\textwidth]{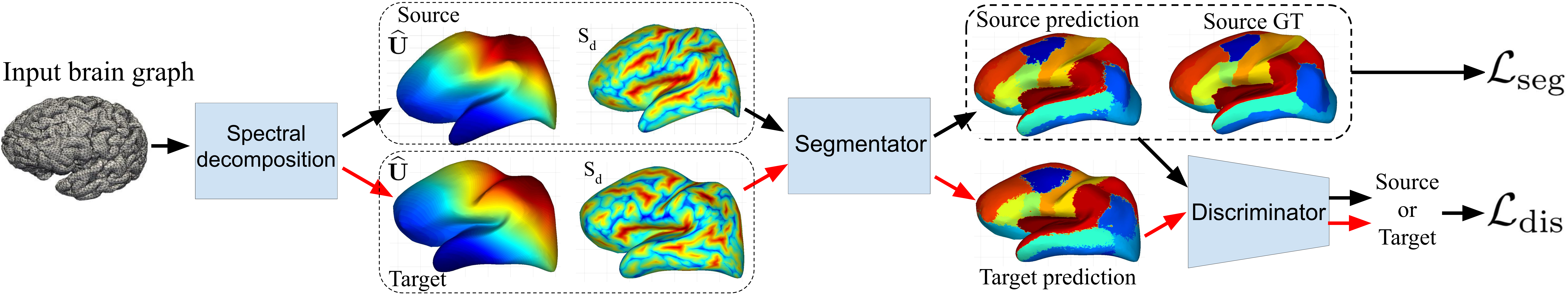}
    %{images/pooling_stg.pdf}
    % if you want to edit the figure: link: 
    % https://docs.google.com/drawings/d/1BXricwNqsBHkb0DCX8XvYf7vWz1dMBGQb-KwvMzE1sM/edit?usp=sharing
    
  \caption{\textbf{Overview of the our architecture:} The input brain graph is mapped to a spectral domain by decomposition of the graph Laplacian. The source and target domain are obtained by aligning the eigenbases to source refernce and targets reference respectively. A segmentator GCN learns to predict a generic cortical parcel label for each domain. The discriminator aims at classifying the segmentator predictions, thereby, assisting the segmentator GCN to adapt to both source and target domains.}
  \label{adv_arch}
\end{figure*}

An overview of our proposed method is shown in Fig. \ref{adv_arch}. In the initial step, the cortical brain graph is embedded into the spectral domain using the graph Laplacian operator. Next, samples only from the source domain are aligned to a reference template using the Iterative Closest Point (ICP) algorithm. Finally, a graph domain adaptation network is trained to perform alignment-independent parcellation. The segmentator network learns a generic mapping from input features of surface data, for instance, the spectral coordinates and sulcal depth of cortical points, to cortical parcel labels. 
\subsection{Spectral embedding of brain graphs}

We start by describing the spectral graph convolution model used in this work. Denote as $\GG = \{\VV, \EE\}$ a brain surface graph with node set $\VV$, such that $|\VV|=N$, and edge set $\EE$. Each node $i$ has a feature vector $\xx_i \in \Real^3$ representing its 3D coordinates. We map $\GG$ to a low-dimension manifold using the normalized graph Laplacian operator $\LL = \II - \DD^{-\frac{1}{2}}\AAA\DD^{-\frac{1}{2}}$, where $\AAA$ is the weighted adjacency matrix and $\DD$ the diagonal degree matrix. Here, we consider weighted edges and measure the weight between two adjacent nodes as the inverse of their Euclidean distance, i.e. $a_{ij} = (\|\xx_i - \xx_j\| + \epsilon)^{-1}$ where $\epsilon$ is a small positive constant. Letting $\LL = \UU \LLambda \tr{\UU}$ be the eigendecomposition of $\LL$, the normalized spectral coordinates of nodes are given by $\Uhat = \LLambda^{-\frac{1}{2}} \UU$. 

Denote the neighbors of node $i \in \VV$ as $\NN_i = \{j \, | \, (i,j) \in \EE\}$. The convolution operation used in our spectral GCN is defined as
\begin{equation}
\begin{aligned}\label{eq:convolution}
    z_{ip}^{(l)} & \ = \
        \sum_{j \in \NN_i} \sum_{q=1}^{M_l} \sum_{k=1}^{K_l} 
            w_{pqk}^{(l)} \, y_{jq}^{(l)} \, \varphi(\uhat_i, \uhat_j; \, \TTheta^{(l)}_k)  \ + \ b_p^{(l)},\\
     y_{ip}^{(l+1)} & \ = \ \sigma\big(z_{ip}^{(l)}\big)
\end{aligned}
\end{equation}
where $y_{jq}^{(l)}$ is the feature of node $j$ in the $q$-th feature map of layer $l$, $w_{pqk}^{(l)}$ is the weight in the $k$-th convolution filter between feature maps $q$ and $p$ of subsequent layers, $b_p^{(l)}$ is the bias of feature map $p$ at layer $l$, and $\sigma$ is a non-linear activation function. The information of the spectral embedding relating nodes $i$ and $j$ is included via a symmetric kernel $\varphi(\uhat_i, \uhat_j; \TTheta_k)$ parameterized by $\TTheta_k$. In this work, we follow \cite{gopinath2019graph} and use a Gaussian kernel: $\varphi(\uhat_i, \uhat_j; \mmu_k, \sigma_k) \ = \ \exp\big(-\sigma_k \, \|(\uhat_j - \uhat_i) - \mmu_k\|^2\big)$.

\subsection{Graph domain adaptation}

Our graph domain adaptation algorithm contains two blocks: a segmentator GCN $\SSS$ performing cortical parcellation and a discriminator GCN $\DDD$ which predicts a given parcellation comes from a source or target graph. Let $\dataSrc$ be the set of source graphs and $\dataTgt$ the set of unlabeled domain graphs, with $\dataSrc \cup \dataTgt$ the entire set of graphs available in training. In the first step, we optimize the segmentator GCN using labeled source graphs $\GG \in \dataSrc$. We feed the segmentation prediction's $\SSS(\GG)$ to the discriminator $\DDD$ whose role is to identify the input's domain (i.e., source or target). The gradients computed from an adversarial loss on target domain graphs is back-propagated from $\DDD$ to $\SSS$, forcing the segmentation to be similar for both the source and target domain graphs. 

As in other adversarial approaches, we define the learning task as a minimax problem between the segmentator and discriminator networks,
\begin{align}\label{eq:loss}
    &\max_{D} \, \min_{S} \ \mathcal{L}(D, S) \, = \, \frac{1}{|\dataSrc|}\sum_{\GG \in \dataSrc} \!\!\!\!\Lseg(S(\GG),\yy_\GG) \ - \ \frac{\lambda}{|\data|}\sum_{\GG \in \data}\!\!\Ldis\big(D(S(\GG)),z_\GG\big),    
\end{align}
where $\Lseg$ is the supervised segmentation loss on labeled source graphs, and $\Ldis$ is the discriminator loss on both source and target graphs, which is optimized in an adversarial manner for $S$ and $D$.

%cross-entropy with Dice on input graph samples and labels from source domain and $\Ladv$ is the weighted cross-entropy adversarial loss that generalizes the segmentation prediction of target graphs to the segmentation of source graphs.

\textbf{Segmentator loss~} For each input graph, the segmentator network outputs a parcellation prediction $\ypred$ where $\widehat{y}_{ic}$ is the probability that node $i$ belongs to parcel $c$. In this work, we define the supervised segmentation loss as a combination of weighted Dice loss and weighted cross-entropy (CE), 
\begin{equation}\label{eq:cross-entropy}
    \Lseg(\ypred,\yy) \, = \, 1 \, - \, \frac{\epsilon \, + \, 2 \sum_{i=1}^N \sum_{c=1}^C \omega_c \, y_{ic} \, \widehat{y}_{ic}}{\epsilon \, + \, \sum_{i=1}^N \sum_{c=1}^C \omega_c (y_{ic} +  \widehat{y}_{ic})} \, - \, \sum_{i=1}^N \sum_{c=1}^C \omega_c \, y_{ic} \, \widehat{y}_{ic}, 
\end{equation}    
with $y_{ic}$ being a one-hot encoding of the reference segmentation and $\epsilon$ a small constant to avoid zero-division. The weights $\omega_c$ balances the loss for parcels by increasing the importance given to smaller-sized regions. In the loss of Eq. (\ref{eq:cross-entropy}), CE improves overall accuracy of node classification while Dice helps to have structured output for each parcel. 

%Second, for the input graphs from the target domain, the output of the forward pass on $\SSS$ is given as input for the $\DD$. 

%We define the adversarial loss over $\GG_D$ as cross-entropy loss:
%\begin{equation}
   %\Ladv(\GG_D) \ = \ - \log \DDD\big(\SSS(\GG_D; \TTheta_S)\big),
%\end{equation}
%where $\TTheta_S = \{w_{pqk}^{(l)}, \,b_p^{(l)}, \,\TTheta^{(l)}_k\}$ are the trainable parameters of the segmentator GCN.

\textbf{Discriminator loss~} Since the discriminator $D$ is a domain classifier, we define its loss as the binary cross-entropy between its domain prediction (i.e., $\zpred=1$ for source or $\zpred=0$ for target):
\begin{equation}
   \Ldis(\zpred, z) \ = \ - \, (1 - z) \log (1-\zpred) \, - \, z \log \zpred.
\end{equation}
As mentioned before, this loss is maximized while updating the segmentator's parameters and minimized when updating the discriminator. Thus, the segmentator learns to produce surface parcellations that are domain-invariant.

\subsection{Network architecture}

We now define the architecture of both the segmentator and discriminator GCN.

\textbf{Segmentator:} The segmentator GCN network is a fully-convolutional network comprised of 3 graph convolution layers with respective feature map sizes of 256, 128, and 32. At the input, each node has 4 features: 3 spectral coordinates and an additional scalar measuring sulcal depth. All layer have $K_l=6$ Gaussian kernels similar to \cite{gopinath2019graph}. Since the output has 32 parcels, our last layer size is set to 32. In the last layer, softmax operation is applied for parcellation prediction, and the remaining layers employ Leaky ReLU as activation function to obtain filter responses in Eq. (\ref{eq:convolution}).

\textbf{Discriminator:} Similar to the segmentator network, we use 2 graph convolution layers, an average pooling layer and 3 fully connected (linear) layers for classifying the segmentation domain. The first graph convolution layer takes a segmentation predictions with 32 feature maps as input. Moreover, the output sizes of the first two layers output are 128 and 64, respectively. Average pooling is used to reduce the input graph to a 1-D vector for the classification task. Three fully-connected layers are placed at the end of the network, with respective sizes of 32, 16 and 1. Each graph convolution layer has $K_l=6$ Gaussian kernels, as in \cite{gopinath2019graph}. Sigmoid activation is applied to the last linear layer to predict the input domain of the graph sample and the remaining layers use Leaky ReLU. 

\vspace{-.5em}
\section{Results} % either "experiments" or "results", not both
\vspace{-.5em}

\begin{figure*}[tb!]
    \centering    
 % \hspace{-15mm}
 \mbox{
    \includegraphics[width=0.42\textwidth]{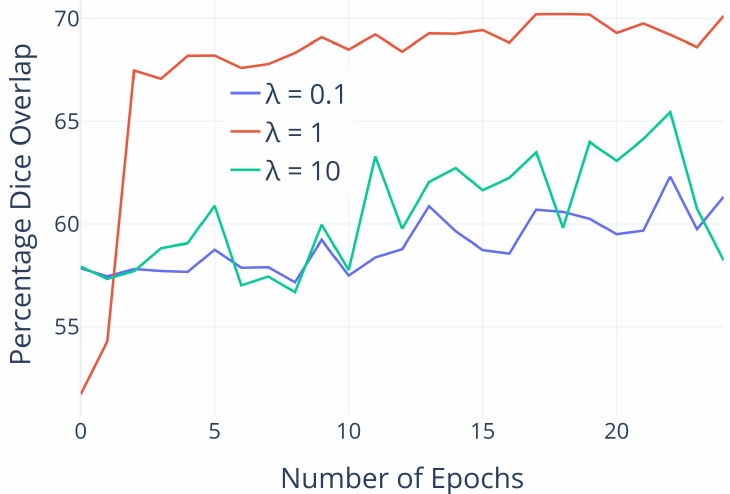}
    
    \hspace{0.5cm}
    
    \includegraphics[width=0.42\textwidth]{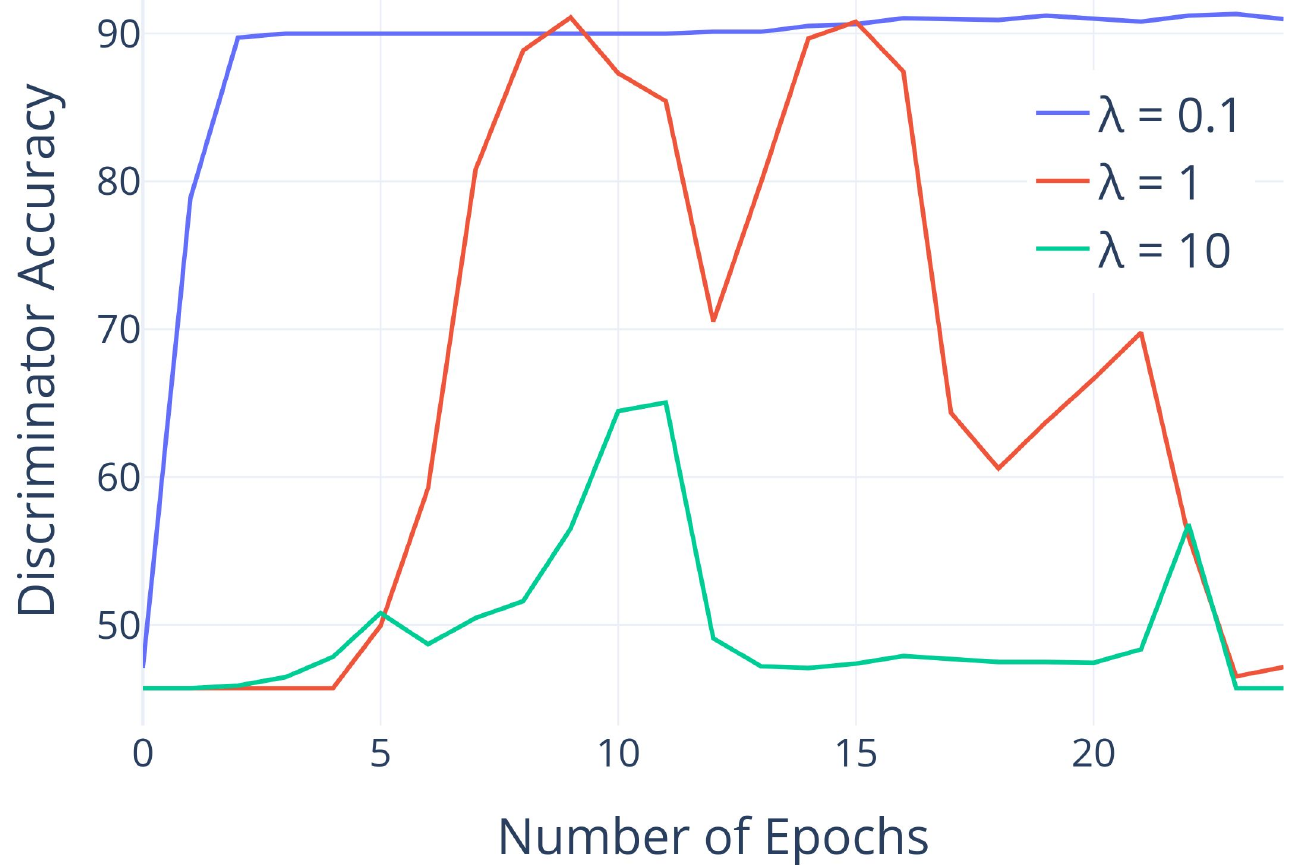}
    }
    \caption{\textbf{Effect of $\lambda$:} The plots shows the segmentator and discriminator performance for different values of $\lambda$. In the \textit{left} plot, we see a significant improvement in performance for $\lambda$ = 1 over $\lambda$ = 0.1 \& 10. The \textit{right} plot shows the discriminator's classification accuracy. The segmentator is unable to fool the discriminator with high value of $\lambda$}
    \label{diff_lam}
    \vspace{-3mm}
\end{figure*}

We evaluate the performance our method using MindBoggle \cite{Klein2017Mindboggling}, the largest manually labelled brain surface dataset. This dataset contains the cortical mesh data of 101 subjects aggregated from multiple sites. Each brain surface includes 32 manually labeled parcels. For each subject, we sub-sample the mesh into 25 smaller sub-graphs with 10k nodes each. All experiments are carried out using this reduced graph on an i7 desktop computer with 16GB of RAM and a Nvidia Titan X GPU. 
First, we assess the impact on segmentation performance of parameter $\lambda$ which controls the relative importance of the supervised segmentation loss and adversarial loss in Eq. (\ref{eq:loss}). Second, we benchmark our domain adaptation algorithm against other learning frameworks for cortical parcellation.

\subsection{Effect of $\lambda$ on segmentation}

The loss function for graph adversarial training involves the hyper-parameter $\lambda$, that controls the effect of adversarial loss on training the segmentator GCN network. We measure the parcellation performance of this network on the target domain surfaces and the discriminator accuracy over $25$ epochs. Our aim is to study how the performance varies with different values of $\lambda \in \{0.1, 1, 10\}$. The mean Dice overlap over epochs for different $\lambda$ values is reported in Fig.~\ref{diff_lam} (left(. Furthermore, the right Fig.~\ref{diff_lam} shows the classification accuracy of the discriminator for the same $\lambda$ values. 

Results of this experiment indicate that $\lambda=1$ is the best choice for training the adversarial GCN. When using a too small $\lambda=0.1$, we observe segmentation performance drop over the unseen target domain surfaces, which illustrates that a stronger adversarial learning is required to align the source and target domains. The dissimilarity between the segmentation predicted for source and target graphs is also evident from the high discriminator accuracy in Fig.~\ref{diff_lam} (right), i.e. the discriminator is not fool in this case. On the other hand, when using a too large $\lambda = 10$, the model focuses mostly on fooling the discriminator, leading to a poor segmentation Dice overlap. Based on this analysis, we will use $\lambda=1$ for the rest of our experiments.

\subsection{Comparison with the state-of-the-art}

\begin{table}[tb!]
\centering
\caption{\textbf{Comparison with surface segmentation approaches:} Average percentage dice overlap and standard deviation on test data. Each domain (column) is generated by aligning the eigenbases of the samples to a reference template. Column 2 is alignned to same source domain reference. Column 3 (None) is unaligned and completely ambiguous. The test set for target domains in columns 4-7 are aligned to random reference from test set.}

\label{tab:results}
\begin{small}
\renewcommand{\arraystretch}{1.15} 
\setlength{\tabcolsep}{1.1pt}
\begin{tabular}{l|cccccc}
\toprule
\multirow[b]{2}{*}{\textbf{Method}} & \multicolumn{6}{c}{\textbf{Test data alignment}} \\
\cmidrule(l{5pt}r{5pt}){2-7}
& \textbf{Source} & \textbf{None} & \textbf{Target\,1} & \textbf{Target\,2} & \textbf{Target\,3} & \textbf{Target\,4} \\ 
\midrule\midrule
Spectral RF \cite{Lombaert2015Spectral}~ &  81.9\ppm3.4   &  65.4\ppm9.0 & 60.0\ppm1.8 & 55.3\ppm2.1 & 60.2\ppm4.0 & 55.2\ppm3.0    \\ 
Seg-GCN~       &  \textbf{86.5\ppm2.8}   &  71.4\ppm7.9 & 67.8\ppm2.0 & 58.8\ppm2.8 & 63.5\ppm3.2 & 60.1\ppm3.6    \\ 
Adv-GCN (ours)~ & 85.7\ppm3.5   &  \textbf{73.8\ppm6.0} & \textbf{73.5\ppm2.0} & \textbf{71.8\ppm2.6} & \textbf{71.0\ppm2.8} & \textbf{71.7\ppm3.3}   \\  
 \bottomrule
\end{tabular}
\end{small}
\end{table}

\begin{figure}[tb!]
    \centering    
    \includegraphics[width=0.95\textwidth]{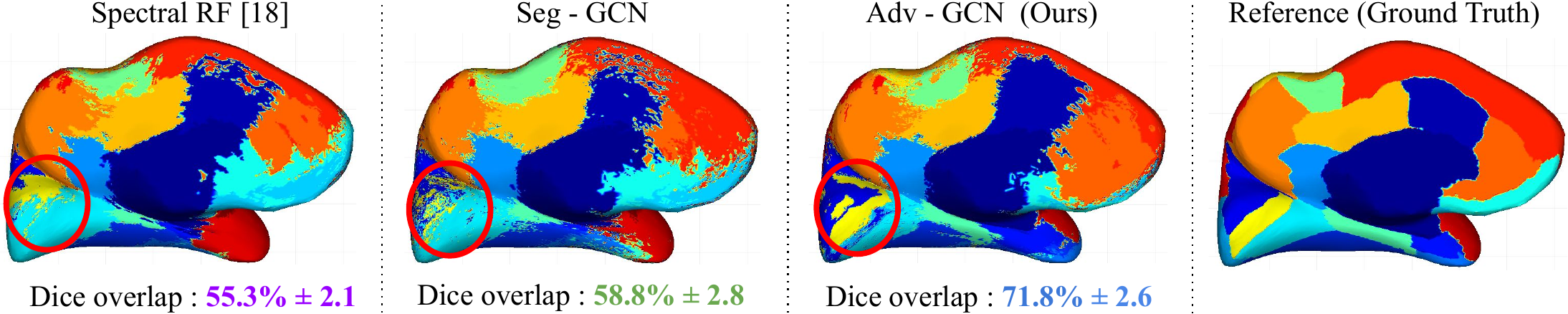}
    \caption{\textbf{Surface segmentation:} The qualitative comparison of different surface segmentation approches with average Dice overlap on target domain. Spectral Ranfom forest \cite{Lombaert2015Spectral} with no neighborhood information has a dice overlap of 55.3$\%$. Seg-GCN considers neigborhood information with a dice overlap of 58.8$\%$. Our graph domain adaptation network, Adv-GCN is able to generalize on target domain (indicated by red circle) with dice overlap of 71.8$\%$. }
    \label{diff_sot}
    % \vspace{-2.4mm}
\end{figure}

We now compare our method with other surface parcellation approaches based on graphs. The average Dice overlap is measured to assess the performance of each model. In Table \ref{tab:results}, we report the performance on unseen test dataset. The different target domains are generated by aligning the eigenbases of test brain graphs either to the same template as source (column 2 -- Source) or completely ambiguous (column 3 -- None) or eigenbases of a random brain graphs from the test set (columns 4 to 7 -- Target 1 to 4).
First, we show the limitation of point-based approaches which ignore the relationship between nodes when predicting labels. Toward this goal, we follow the spectral random forest (RF) approach in \cite{Lombaert2015Spectral} and train a random forest with 50 trees using the same input as given to GCN networks (i.e., spectral coordinates and sulcul depth). As shown in Table \ref{tab:results}, this Spectral RF approach achieves a mean Dice overlap of 81.9$\%$ with the aligned set and only 65.4$\%$ on unaligned set. The random forest does not consider neighborhood information for parcellation and thus obtains low performance on the unaligned brain graphs. A graph segmentation network without additional discriminator (Seg-GCN) network yield an average percentage Dice overlap of 86.5$\%$ on aligned set and 71.4$\%$ on unaligned set. We achieve an improvement in performance of 4.6$\%$ and 6.0$\%$ on aligned and unaligned domain respectively with only additional neighborhood information used by GCN segmentation network. Further, our GCN network trained in an adversarial setting (Adv-GCN) produces generic segmentation maps on both aligned and unaligned brain graphs. An average percentage Dice overlap of 85.7$\%$ on aligned set and \textbf{73.8$\%$} on unaligned set. Our proposed model Adv-GCN has an increased in performance over unaligned set with equivalent performance on aligned set to Seg-GCN. To better understand the significance of our graph domain adaptation network, we evaluate our method against multiple aligned domains. The Table  shows the our model achieves an improved performance across unaligned domain and different target aligned domains. The Figure \ref{diff_sot} shows qualitative results for different graph segmentation methods.

%dm-ori
%70.6004
%2.8378
%78.3468
%2.9401

% \begin{table}[tb!]
% \centering
% \caption{Percentage Dice on test set}
% \label{tab:results}
% \begin{small}
% \setlength{\tabcolsep}{2pt}
% \begin{tabular}{lccccccc}
% \toprule
% \multirow[b]{2}{*}{\textbf{Method}} & \multirow[b]{2}{*}{\textbf{Source}} & \multicolumn{6}{c}{\textbf{Target alignment}} \\
% \cmidrule(l{5pt}r{5pt}){3-8}
% & & None & Target 1 & Target 2 & Target 3 & Target 4 & Target 5 \\ 
% \midrule\midrule
% RF            &  89.2\ppm4.5   &  72.0\ppm3.8 & 70.0\ppm3.6 & 73.4\ppm7.6 & 63.1\ppm4.0 & 66.0\ppm2.8 & 64.4\ppm3.8   \\ 
% Seg-GCN~       &  90.7\ppm2.7   &  79.5\ppm8.4 & 75.8\ppm3.2 & 76.2\ppm2.7 & 70.4\ppm3.1 & 76.2\ppm4.4 & 70.2\ppm4.1   \\ 
% Adv-GCN~       &  Avg 90.1       &  83.2\ppm6.5 & 86.8\ppm1.6 & 85.8\ppm3.4 & 86.1\ppm4.1 & 86.1\ppm3.9 & 84.8\ppm3.7   \\  

%  \bottomrule
% \end{tabular}
% \end{small}
% \end{table}

\section{Conclusion}

In this paper, we present a novel adversarial domain adaptation framework for brain surface graphs. The proposed algorithm leverages a adversarial training mechanism to obtain a generalized brain surface segmentation. The reported experiments illustrate the advantages of our approach for brain surface segmentation. This method overcomes the limitations of spectral GCNs \cite{bruna2013spectral,Defferrard2016Convolutional} that require finding an explicit alignment of graph eigenbases. The Table \ref{tab:results} shows a clear improvement in performance over the latest spectral GCN \cite{bruna2013spectral,Defferrard2016Convolutional} as well as the forest-based \cite{Lombaert2015Spectral} approaches. Our method improves the average Dice performance for parcellation by \textbf{$2.4\%$} over unaligned domains and a maximum of different over multiple domain alignments. The performance and time complexity of our method is similar to Seg-GCN \cite{gopinath2019graph} on test sets for a source domain. The Fig. 3 illustrates the qualitative comparison of our adversarial GCN. The potential of our adversarial graph domain adaptation technique is demonstrated on surface segmentation, but can also be used for other surface segmentation problems. For example, domain adaptations for semi-supervised segmentation, thereby mitigating the requirement of large amounts of labelled surfaces. 

%
% ---- Bibliography ----
%

% \herve{can we have a little bit more refs?}

\subsubsection*{Acknowledgments}
This research work was partly funded by the Fonds de Recherche du Quebec (FQRNT) and Natural Sciences and Engineering Research Council of Canada (NSERC). We gratefully acknowledge the support of NVIDIA Corporation for the donation of the Titan X Pascal GPU used for this research.

\bibliographystyle{splncs}
\bibliography{Reference}

\end{document}